\documentclass[11pt]{article}

\usepackage{amsmath,amssymb}

\usepackage{epsfig}
\usepackage{graphicx}               
\usepackage{url}
\usepackage{hyperref}
\usepackage{cite}
\usepackage{color}

\newcommand{\nc}{\newcommand}

\nc{\beq}{\begin{equation}}
\nc{\eeq}{\end{equation}}
\nc{\beqa}{\begin{eqnarray}}
\nc{\eeqa}{\end{eqnarray}}
\nc{\bea}{\begin{eqnarray}}
\nc{\eea}{\end{eqnarray}}
\nc{\ra}{\rightarrow}
\nc{\lsim}{\begin{array}{c}\,\sim\vspace{-21pt}\\< \end{array}}
\nc{\gsim}{\begin{array}{c}\sim\vspace{-21pt}\\> \end{array}}
\nc{\Tr}{{\rm Tr}}
\nc{\slsh}{\slash\hspace*{-0.22cm}}

\def\be{\begin{equation}}
\def\ee{\end{equation}}
\def\bea{\begin{eqnarray}}
\def\eea{\end{eqnarray}}
\def\bit{\begin{itemize}}
\def\eit{\end{itemize}}

\def\to{\rightarrow}

\title{
\vspace*{-2.3cm}
\begin{flushright}
\normalsize{
  }
\end{flushright}
\vspace{1.5cm}
\Large
\textbf{
Dips at Colliders
}\vspace*{1.0cm}
}

\author{Yang Bai\,$^{a}$ and Wai-Yee Keung\,$^{b}$
\vspace{5mm}
\\
$^{a}$ \normalsize\emph{Department of Physics, University of Wisconsin, Madison, WI 53706, USA}  \vspace{1mm} \\
$^{b}$ \normalsize\emph{Physics Department, University of Illinois at Chicago, Chicago IL 60607, USA}
}

\date{}

\begin{document}
\setcounter{page}{0}
\maketitle

\vspace*{1cm}
\begin{abstract}
We categorize new physics signatures that manifest themselves as a ``dip" structure at colliders. One potential way to realize a dip is to require interactions to be zero when all particles are mass on-shell, but not if one or more are mass off-shell. For three particle interactions, we have found three interesting cases: one massive gauge boson with two identical scalars; one massless gauge boson with two different scalars; one massive gauge boson with two identical massless gauge bosons. For each case, we identify the relevant effective operators to explore its dip signature at the LHC. Unfortunately, the unstable particle with a vanishing mass-on-shell interaction has a complex mass which is coincident with the complex pole in its propagator. As a result, a contact-like amplitude without a dip is produced. We then point out two other interesting ways that generate a dip in the cross section. The first way has a dip signature due to a zero in the vertex form factor of the time-like momentum in the $s$-channel. In the second way, the dip plus bump signature appears because there is destructive interference among processes of exchanging different $s$-channel particles.
\end{abstract}

\thispagestyle{empty}
\newpage

\setcounter{page}{1}

\baselineskip18pt

\vspace{-3cm}
\section{Introduction}
\label{sec:intro}
The Large Hadron Collider (LHC) has achieved a great triumph from its discovery of the Higgs boson~\cite{Aad:2012tfa,Chatrchyan:2012ufa} in 2012 at 8 TeV center-of-mass energy. The next big goal of the LHC running with a higher center-of-mass energy is to discover new physics beyond the Standard Model (SM). With hundreds of experimental searches with many combinations of objects, there is still no sign of new physics. One simple explanation is that the LHC has not reached its full colliding energy as well as luminosity. Some new particles may evade the current searches because it is too heavy or buried in the SM background. We could be more patient and wait for the LHC Run 2 to tell us the physics at the TeV scale. On the other hand, one may wonder whether the existing search strategy at the LHC is drawing water with a sieve. Some non-trivial signatures may require non-standard search strategies and deserves us attention to identify and search for them. 

A big fraction of existing LHC searches have been concentrated on resonances or ``bump hunting".  For instance, the Higgs boson as a di-photon or four-lepton resonance is one of the good examples. The search for resonances in models beyond the SM will continue to be the major task in the high energy frontier. However, sometimes due to some selection rule, new physics may not show up as a resonance, but instead has a sign of a suppressed event rate around the new particle mass. The new particle may then manifest itself as a ``{\it dip}" in terms of invariant masses of some objects. In this paper, we try to systematically identify the cases for a new particle to behave as a dip at colliders. 

The easiest way to obtain a dip in kinematic distributions is to have interference of two or more particles. Even in the SM, the $e^+e^- \rightarrow \gamma^*/Z \rightarrow \mu^+\mu^-$ has a faint dip before the resonance (the $Z$ boson) location. One could establish the existence of the $Z$ boson at the VENUS experiment of TRISTAN~\cite{Abe:1990rr} with a center-of-mass energy up to 64 GeV after the resonance discovery of the $Z$ boson at the UA-1/2 of CERN SppS in 1983. Similarly, one could have interference among $\gamma^*$, $Z$ and a new $Z^\prime$ to have a dip structure in the di-muon or di-tau invariant mass distribution before the bump at the $Z^\prime$ mass. In this paper, we will provide one example of $Z^\prime$'s with an obvious dip plus bump structure. 

A more interesting way is how to obtain only a dip structure without a bump, which will be the focus of our paper.  For three particle interactions of a new particle with two SM particles, one can translate this question as under what conditions one can have a vanishing vertex when all three particle mass on-shell, but a non-vanishing vertex when one or more particles mass off-shell. One of the famous examples is the Landau-Yang theorem~\cite{Landau:1948kw,Yang:1950rg}, which states that one massive spin-one particle can not decay into two identical massless spin-one particles. This theorem means that when all three spin-one particles are mass on-shell, the interaction vanishes. However, when one of the three particles is off-shell, the interaction is not zero. 

To identify more cases satisfying the condition to provide a dip, 
we will use both spinor-helicity method (in Section~\ref{subsec:spinor-helicity}) and operator analysis to systematically cover all cases with particle spins below two. For three particle interactions, we have found three interesting cases for providing a mass-on-shell interaction: one massive gauge boson with two identical scalars; one massless gauge boson with two different scalars; one massive gauge boson with two identical massless gauge bosons. We will first use the spinor-helicity method to simply show that those three cases satisfy vanishing mass-on-shell but non-vanishing mass-off-shell matrix element. 

Then in Section~\ref{subsec:mass-on-shell-dip}, 
for every case we introduce one effective operator to
illustrate whether the dip structure shows up at the LHC or not. We point out the subtleties of choosing the correct massive gauge boson propagators associated with an unstable particle. For an unstable particle, a complex mass should be used to define its mass on-shell condition. The same complex mass also shows up in this unstable massive particle propagator. As a result, the zero in the numerator and the zero in the denominator cancel and leave a contact-like interaction without a dip.

For the case related to the Landau-Yang theorem in Section~\ref{sec:case-III}, we point out the interesting collider signatures for the fermion-phobic $Z^\prime$. We provide a detailed analysis of $Z^\prime$ productions from 
gluon-gluon fusion. The amplitudes in $gg\to
gg$ due to the new $Z^\prime$ as well as its interference with the standard QCD process are evaluated.  If the same $Z^\prime$ couples to the top quark axial-vectorially, the $Z^\prime$ may give rise to
observable excess in the large $m_{t\bar t}$ tail. This is because the effective amplitude resembles a higher-dimensional operator with the top pair.

In Section~\ref{sec:form-factor}, we explore another way to generate a dip structure by constructing a zero in the vertex form factor of the time-like momentum transfer. As a concrete example, we show that a heavy scalar coupling to two gluons can have a zero in the interaction form factor. We use this form factor to demonstrate the existence of the dip signature, although a bigger bump may also appear  at a much higher center-of-mass energy away from the dip location. 

In Section~\ref{sec:dip-bump}, we review the ``standard way'' of realizing a dip that occurs due to
destructive interference between resonances. Such a dip is located in the side band of the new particle when the
off-resonance amplitudes from various channels delicately cancel each
other. Therefore, the dip is always accompanied with a nearby bump
of a resonance.

In this article, we do not address the interesting case of the
radiation amplitude zero in certain processes 
at a specific scattering angle, where the differential cross section vanishes. For instance, one has 
the SM process
$u\bar d \to \gamma W^+$\cite{Mikaelian:1979nr}, 
the high $p_T$ chargino-neutralino production of 
SUSY\cite{Barger:1983wc,Brown:1984nj,Hewett:2011fu}, and the lepto-quark process\cite{Deshpande:1994vf,Doncheski:1998cv}.  Usually the radiation amplitude zero does not maintain zero when going beyond the tree level, and its detection requires extreme angular resolution in the subprocess frame.

\section{Three-particle Interactions with a Vanishing Mass-on-shell Vertex}
\label{sec:dip-structure}
In this section, we first use the spinor-helicity formulas to find all possible cases with a a vanishing mass-on-shell vertex among three particles. We then discuss whether one can have a dip structure based on specific operator examples. 
\subsection{Spinor-helicity Formalism}
\label{subsec:spinor-helicity}
For three particle interactions, we restrict ourselves to spins below two. If all particles are bosons (we have not found interesting examples for fermion interactions), we have combinations of $(0, 0, 0)$, $(0,0,1)$, $(0,1,1)$ and $(1,1,1)$. The cases of $(0,0,0)$ and $(0,1,1)$ (for instance, $h\rightarrow \gamma Z$ and $h\rightarrow \gamma \gamma$) have non-zero matrix elements when all three particles are on-shell. 

For the remaining two cases, $(0,0,1)$ and $(1,1,1)$, we use the
spinor-helicity formalism~\cite{DeCausmaecker:1981bg, Xu:1986xb} to
find the cases with zero matrix elements when all three particles are
on-shell. We follow the notation in the textbook\cite{Zee:2003mt}.
In this formalism, the four-vector momentum $p^\mu$ is replaced by a product of two spinors: $p_{\alpha \dot{\alpha}}\equiv p_\mu (\sigma^\mu)_{\alpha \dot{\alpha}}$. The product of two vectors is $p\cdot q = \epsilon^{\alpha \beta} \epsilon^{\dot{\alpha} \dot{\beta}}p_{\alpha \dot{\alpha}}q_{\beta \dot{\beta}}$. For light-like four momentum, one has $p_{\alpha \dot{\alpha}}= \lambda_\alpha \tilde{\lambda}_{\dot{\alpha}}$ in terms of two spinors. For two light-like momenta, one can introduce the ``angle spinor bracket" and ``square spinor bracket" to simplify the spinor index contraction. For $p_{\alpha \dot{\alpha}}= \lambda_\alpha \tilde{\lambda}_{\dot{\alpha}}$ and $q_{\alpha \dot{\alpha}}= \mu_\alpha \tilde{\mu}_{\dot{\alpha}}$, we have 
\beqa
p\cdot q  = (\epsilon^{\alpha \beta} \lambda_\alpha \mu_\beta) (\epsilon^{\dot{\alpha} \dot{\beta}} \tilde{\lambda}_{\dot{\alpha}} \tilde{\mu}_{\dot{\beta}} )  \equiv \langle \lambda \mu \rangle [ \tilde{\lambda} \tilde{\mu} ]
\equiv \langle \lambda \mu \rangle [ \lambda \mu ]   \,.
\eeqa
For massless photon, one can use the requirement of $\epsilon(p) \cdot p = 0$ to specify the two different polarizations, which are 
\beqa
\epsilon^-_{\alpha \dot \alpha} = \frac{ \lambda_\alpha \tilde{\mu}_{\dot \alpha} }{ [ \lambda \mu]} \,, \qquad
\epsilon^+_{\alpha \dot \alpha} = \frac{ \mu_\alpha \tilde{\lambda}_{\dot \alpha} }{ \langle \mu \lambda \rangle} \,,
\eeqa
for arbitrary $\tilde{\mu}_{\dot \alpha}$ and $\mu_\alpha$, which represent the freedom inherent in a gauge theory. For many massless vectors, $\epsilon_i$, one has 
\beqa
\epsilon^+_i \cdot \epsilon^+_j = \frac{\langle \mu_i \mu_j \rangle[\lambda_i \lambda_j] }{\langle \mu_i \lambda_i \rangle[\mu_j \lambda_j]} \,, \quad
\epsilon^-_i \cdot \epsilon^-_j = \frac{\langle \lambda_i \lambda_j \rangle[\mu_i \mu_j] }{[ \lambda_i \mu_i ]\langle \lambda_j \mu_j \rangle } \,, \quad
\epsilon^-_i \cdot \epsilon^+_j = \frac{\langle \lambda_i \mu_j \rangle[\mu_i \lambda_j] }{[ \lambda_i \mu_i ]\langle \mu_j \lambda_j \rangle }  \,.
\eeqa
If the vertex contains only $n$ massless gauge fields under the same gauge symmetry, one has the conservation of momenta such that one can have the freedom to choose $n-1$ arbitrary $\mu_i$'s. For instance, one can prove that the amplitudes $A(+++\cdots ++)$ and $A(-++\cdots ++)$ are zero by choosing $\mu_1=\mu_2=\cdots = \mu_n$ and $\mu_2=\mu_3\cdots = \mu_n = \lambda_1$, respectively. 

Now, come back to our first case of $(0, 0, 1)$. There are two interesting sub-cases. The first sub-case has a massive spin-1 particle and two identical spin-0 particles. Because the Bose symmetry between the two scalar momenta, $p_1$ and $p_2$, the matrix element should be proportional to $(p_1 + p_2) \cdot \epsilon_3 (p_3)=-p_3 \cdot \epsilon_3 (p_3) =0$, where we have defined all momenta to be out-going from the vertex. For an off-shell massive gauge boson, $p_3 \cdot \epsilon_3 (p_3) \neq 0$ and we can have a non-zero matrix element. The second sub-case has a massless spin-1 particle and two different spin-0 particles. If the massless spin-1 particle is mass on-shell, we have $p_3 = -(p_1 + p_2)$ to be light-like. So, we can write $p_3 = \lambda_3 \tilde{\lambda}_3$ in the spinor notation. There are only two possible contractions for the gauge boson polarization vector: $(p_1 + p_2)\cdot \epsilon_3$ and $(p_1 - p_2)\cdot \epsilon_3$. The first one is obviously zero. The second one can be rewritten as $(p_1 - p_2) \cdot \epsilon^+_3 \propto (p_1 - p_2) \cdot \mu_3 \tilde{\lambda}_3$. If one chooses $\mu_{3\, \alpha} = \epsilon_{\alpha \beta} \lambda_3^\beta$, one can easily show that this matrix element is zero. A similar argument can show a vanishing value for $(p_1 - p_2) \cdot \epsilon^-_3$. On the other hand, if the gauge boson is off-shell, the matrix element is non-zero. 

For the second case of $(1, 1, 1)$ and following the Landau-Yang's theorem, we consider the case of two identical massless gauge bosons with momenta, $p_1$ and $p_2$, plus one massive gauge boson with a momentum $p_3$. If the matrix element does not contain the $\varepsilon^{\alpha \beta \gamma \sigma}$ tensor, we only need to use $\epsilon_1(p_1)$, $\epsilon_2(p_2)$, $\epsilon_3(p_3=-p_1-p_2)$, $p_1$ and $p_2$ to build the matrix element. Requiring the matrix element to be symmetric under $1 \leftrightarrow 2$, we only have two options after using $p_1\cdot \epsilon_1(p_1)=0$, $p_2\cdot \epsilon_2(p_2)=0$ and $p_3\cdot \epsilon_3(p_3)=0$:
\beqa
&& \epsilon_3 \cdot (p_1 - p_2) ( \epsilon_1 \cdot p_2 - \epsilon_2 \cdot p_1 )  \,, \nonumber \\
&& \epsilon_3 \cdot \epsilon_1\, \epsilon_2 \cdot p_1 + \epsilon_3 \cdot \epsilon_2\, \epsilon_1 \cdot p_2 \,.
\label{eq:matrix-element}
\eeqa
For the two massless bosons and noticing that 
\beqa
\epsilon^+_i \cdot p_j = \frac{ \langle \mu_i \lambda_j \rangle [ \lambda_i \lambda_j ] }{\langle \mu_i \lambda_i \rangle } \,, \qquad
\epsilon^-_i \cdot p_j = \frac{ \langle \lambda_i \lambda_j \rangle [ \mu_i \lambda_j ] }{ [ \lambda_i \mu_i ] }
\,,
\eeqa
one can choose $\mu_1 = \lambda_2$ and $\mu_2 = \lambda_1$ and hence $\epsilon_1 \cdot p_2 =\epsilon_2 \cdot p_1 = 0$ to show that both terms in Eq.~(\ref{eq:matrix-element}) are zero. 

For the interactions containing the $\varepsilon^{\alpha \beta \gamma \sigma}$ tensor, there is only one option:
\beqa
\varepsilon_{\alpha \beta \gamma \sigma} \epsilon^\alpha_3 \epsilon^\beta_1 \epsilon^\gamma_2 ( p_1^\sigma  - p_2^\sigma ) \,,
  \label{eq:matrix-element-2}
\eeqa
after the requirement of Bose symmetry of the two massless bosons. For the choice of $\mu_1 = \lambda_2$ and $\mu_2 = \lambda_1$, we have $p_3\cdot \epsilon_{1} = p_3\cdot \epsilon_{2} = p_3\cdot \epsilon_{3}=0$. On the other hand, we have $p_3 \cdot (p_1 - p_2) = p_1^2 - p_2^2 =0$. So, the vector $(p_1 - p_2)$ should be a linear combination of $\epsilon_1$, $\epsilon_2$ and $\epsilon_3$. Therefore, the interaction in Eq.~(\ref{eq:matrix-element-2}) vanishes when all three particles are mass on-shell, but does not vanish if one particle is mass off-shell. The above approach to the Landau-Yang theorem using the helicity method supplements to the standard proof by using the rotation symmetry in Ref.~\cite{Landau:1948kw,Yang:1950rg,Keung:2008ve}.

In summary, we have found three cases with vanishing on-shell matrix elements but non-zero off-shell matrix elements, as summarized in Table~\ref{table:three-case-summary}.
\begin{table}[htb!]
\renewcommand{\arraystretch}{1.5}
\begin{center}
\begin{tabular}{c|l}
\hline \hline
Case I & \mbox{1 massive gauge boson + 2 identical scalars} \\ \hline
Case II & \mbox{1 massless gauge boson + 2 different scalars}  \\ \hline
Case III & \mbox{1 massive gauge boson + 2 identical massless gauge bosons}  \\ \hline
\end{tabular} 
\caption{Three cases with zero mass-on-shell matrix elements but non-zero mass-off-shell matrix elements.}
\label{table:three-case-summary}
\end{center}
\end{table}
\,
%

\subsection{Can we have Dip Signatures at the LHC for the above Three Cases?}
\label{subsec:mass-on-shell-dip}
Naively speaking, the vanish of the interactions when the intermediate particles stay mass on-shell could lead to an obvious dip signatures at colliders. This will be true if the denominator of the intermediate particle propagator does not vanish at the same mass-on-shell condition. In this section, we write down the representative effective operators for all three cases and study their manifestation at the LHC. Similar studies can easily be performed at a linear collider like ILC, CEPC or TLEP. We will point out the subtleties of some cases where there are always off-shell contributions to the matrix element such that a dip structure signature does not appear. 

\subsubsection{Case I: one massive gauge boson with two identical scalars}
\label{sec:case-I}
For this case, we choose the massive gauge boson to be a new $Z^\prime$ and the two identical scalars to be the Higgs boson in the SM. The gauge invariant dimension-six operator is
\beqa
{\cal O}_{Z^\prime hh} = \frac{i\,(H^\dagger D_\mu H)(S^\dagger D^\mu S)}{\Lambda_1^2} \,+\, h.c. \,,
\eeqa
with $\langle S \rangle = v^\prime/\sqrt{2}$ to spontaneously break the $U(1)^\prime$ symmetry. After the electroweak symmetry breaking with $H^T = [0, (v + h)/\sqrt{2}]$, we have 
\beqa
{\cal O}_{Z^\prime hh} \supset  - \frac{g^\prime \, v^{\prime\, 2} }{2\,\Lambda_1^2} h\, \partial_\mu h \, Z^{\prime\, \mu} = \frac{g^\prime \, v^{\prime\, 2} }{4\,\Lambda_1^2} h^2 \, \partial_\mu \,Z^{\prime\, \mu} = \frac{M^2_{Z^\prime}}{4\, g^\prime \Lambda_1^2} \,h^2 \, \partial_\mu \,Z^{\prime\, \mu}   \,,
\eeqa
with $g^\prime$ as the gauge coupling of $U(1)^\prime$ and the charge of $S$ as one under $U(1)^\prime$. It is obvious that when $Z^\prime$ is mass on-shell the interaction vanishes. 

\begin{figure}[th!]
\begin{center}
\includegraphics[width=0.6\textwidth]{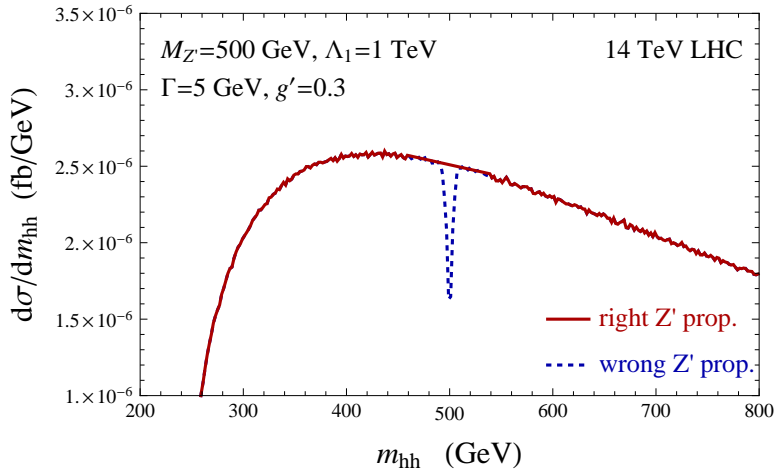}
\caption{The differential cross section of an off-shell $Z^\prime$ production in terms of the invariant mass of two Higgs bosons: $pp \rightarrow h\,Z^{\prime *} \rightarrow h\,h\,h$. The dip structure in the blue and dotted curve has used a wrong $Z^\prime$ propagator. {\it From a gauge-invariant calculation, there is no dip in the $m_{hh}$ spectrum, as shown in the red and solid curve.}}
\label{fig:doublehiggs}
\end{center}
\end{figure}

The signatures at the LHC for this specific model are from the process of $gg \rightarrow h \, Z^{\prime *} \rightarrow h\,h\,h$. There are three SM Higgs bosons in the final state, among them two Higgs bosons have their invariant mass spectrum with a potential dip structure. We implement this new $Z^\prime$ model in \texttt{FeynRules}~\cite{Alloul:2013bka} and use \texttt{CalcHEP}~\cite{Belyaev:2012qa} to generate kinematic distributions. In the blue and dotted curve of Fig.~\ref{fig:doublehiggs}, we show the simulated distributions of two Higgs boson invariant masses at the 14 TeV LHC for a 5 GeV $Z^\prime$ width. Even after the dilution of three combinations of pairing, one still can see a clear dip in the invariant mass distributions. On the other hand, due to the three-body final state, the signal production cross section is tiny and requires a high luminosity LHC to dig it out. 

However, there is a flaw for the results obtained in the blue and dotted line of Fig.~\ref{fig:doublehiggs}. The mistake comes from how to properly write down the $Z^\prime$ propagator when the $Z^\prime$ has a width. In the unitary gauge, the correct and gauge-invariant propagator for the $Z^\prime$ should be
\beqa
\frac{-i}{q^2 - (M_{Z^\prime}^2 - i \Gamma_{Z^\prime} M_{Z^\prime}) } \left( g^{\mu\nu} - \frac{q^\mu q^\nu}{M_{Z^\prime}^2 - i \Gamma_{Z^\prime} M_{Z^\prime} } \right)  \,.
\eeqa
The numerical calculations to generate the blue and dotted line in Fig.~\ref{fig:doublehiggs} has neglected the width part in the numerator of the above propagator formula. As studied for the SM $Z$ boson properties in Refs.~\cite{Stuart:1991xk,Willenbrock:1991hu,quant-ph/0312178,Artoisenet:2013puc}, one should do the replacement of $M_{Z^\prime}^2 \rightarrow  M_{Z^\prime}^2 - i \Gamma_{Z^\prime} M_{Z^\prime}$ to have gauge invariant results. After multiplying $q^\mu q^\nu$  from the two vertexes of $Z^\prime h h$, we have the matrix element proportional to $q^2/(M_{Z^\prime}^2 - i \Gamma_{Z^\prime} M_{Z^\prime})$, which does not show a dip structure. The correct distribution without a dip structure is shown in the red and solid line of Fig.~\ref{fig:doublehiggs}. On the other hand, if one only keeps the width term in the denominator and neglects the part in the numerator, one has a wrong matrix element proportional to $q^2\,(q^2 - M_{Z^\prime}^2)/[M_{Z^\prime}^2(q^2 - M_{Z^\prime}^2 + i \Gamma_{Z^\prime} M_{Z^\prime})]$, which vanishes when $q^2 = M_{Z^\prime}^2$.

The vanishing interaction for a mass-on-shell $Z^\prime$ can be seen from $q\cdot \epsilon(q) = 0$, which demands only three degrees of freedom for a mass-on-shell $Z^\prime$. So, if $Z^\prime$ is a stable particle with a real mass, we anticipate vanishing interactions or even vanishing matrix element when $q^2 = M_{Z^\prime}^2$. The story changes when $Z^\prime$ is an unstable particle. One needs to introduce a complex mass to define the mass-on-shell condition. For the case at hand, we have the complex mass to be $\widetilde{M}_{Z^\prime}^2 \equiv M_{Z^\prime}^2 - i \Gamma_{Z^\prime} M_{Z^\prime}$. To satisfy the mass-on-shell condition, one needs to have a complex $q^2$ with $q^2 = \widetilde{M}_{Z^\prime}^2$, which can not be realized in a realistic experiment. Even if a complex mass can be probed, the cancellation of an identical root in the numerator and denominator still generates a non-vanishing matrix element. 

There is another way to see that there is no dip structure for the operator in ${\cal O}_{Z^\prime h h }$. From spontaneous gauge symmetry breaking, one generates a term $G_{Z^\prime} \,\partial^\mu Z^\prime_\mu$ with $G_{Z^\prime}$ as the Goldstone boson field. One then chooses a gauge fixing term to get rid of this term. For our case, one could also cancel the $h^2 \partial^\mu Z^\prime_\mu$ term at the same time when we choose a gauge-fixing term, for instance, a non-linear gauge-fixing term~\cite{Fujikawa:1973qs}. After doing so, only the Goldstone boson $G_{Z^\prime}$ couples to two $h$'s plus a gauge-parameter-dependent quartic contact interaction of $h^4$. The summation of matrix elements from the $G_{Z^\prime}$-mediated $h h \rightarrow  h  h$ process and from the quartic $h^4$ interaction render a gauge-invariant result proportional to $m_{hh}^2$ and has no dip structure.


\subsubsection{Case II: one massless gauge boson with two different scalars}
\label{sec:case-II}
For this case, the simplest demonstration is a complex scalar, $S\equiv (S_1 + i\,S_2)/\sqrt{2}$, with a charge radius operator under the electromagnetic interaction. The effective operator is
\beqa
{\cal O}_{\gamma S S^\dagger}  = \frac{i\,e\, \partial_\mu S^\dagger \partial_\nu S \, F^{\mu\nu} } {2\,\Lambda_2^2} \,+\, h.c. = \frac{e\,\partial_\mu S_2\, \partial_\nu S_1 \, F^{\mu\nu} } { \Lambda_2^2}  \,.
\eeqa
Using integration by parts, it is easily to show that this operator becomes $(S_2 \partial_\nu S_1) J^\nu$, so the two real scalars only couple to the electromagnetic current or an off-shell photon. The two real scalars, $S_1$ and $S_2$, could have different masses, $M_{S_1}$ and $M_{S_2}$, respectively. Without loss of generality, we will assume $M_{S_1}< M_{S_2}$ such that $S_2$ can decay into $S_1$ plus an off-shell photon. Depending on the mass difference, one could have $S_2 \rightarrow S_1 \,\gamma^* \rightarrow S_1 \, \ell^- \ell^+\,(S_1 \,jj, S_1 \,t\bar{t})$. The lighter scalar $S_1$ could be a stable particle at colliders if it has no large coupling to the SM particles. As an illustration, we don't introduce additional operator to make $S_1$ decay and consider it as a stable particle at colliders.

\begin{figure}[th!]
\begin{center}
\includegraphics[width=0.45\textwidth]{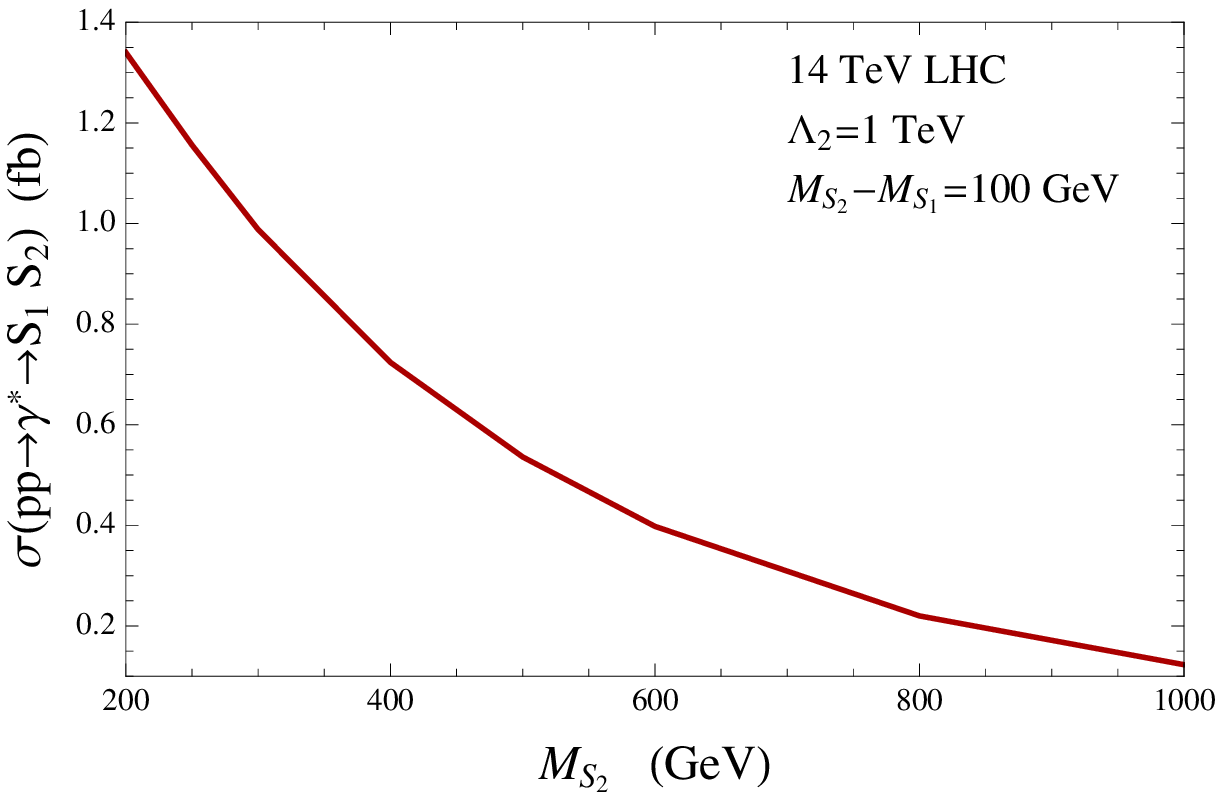} \hspace{3mm}
\includegraphics[width=0.465\textwidth]{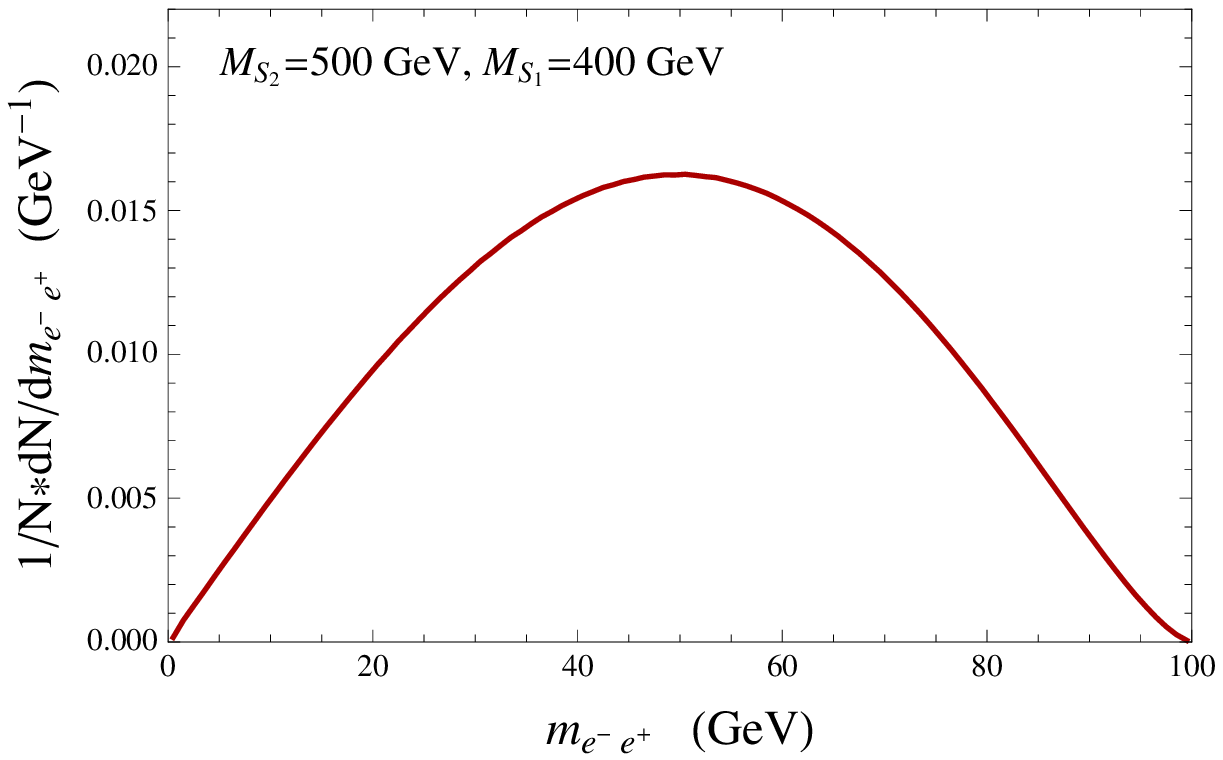}
\caption{Left panel: the production cross section of $pp \rightarrow \gamma^* \rightarrow S_1 S_2$ at the 14 TeV LHC. Right panel: the distribution of invariant masses of electron and positron from $S_2 \rightarrow S_1\, \gamma^* \rightarrow S_1\,e^- \,e^+$.}
\label{fig:doublelectron}
\end{center}
\end{figure}

At the LHC, one has the production of $p p \rightarrow \gamma^* \rightarrow S_1 \,S_2$ with $S_2  \rightarrow S_1 \, \ell^- \ell^+$ as an example. So, the final signature is $\ell^- \ell^+ + \mbox{MET}$. As one can already guess, although the final state of this model matches to that of pair-productions of SM weak gauge bosons and sleptons in the Minimal Supersymmetric Standard Model (MSSM), the invariant mass distributions of the two leptons could be dramatically different. The production cross section of $u\bar{u} \rightarrow \gamma^* \rightarrow S_1 \,S_2$ as a function of the parton center-of-mass energy is calculated to be
\beqa
\hat\sigma(\hat s) = \frac{e^4\,Q_u^2}{576\,\pi\,\Lambda_2^4\,\hat{s}^2}\,\left[ \left(\hat s - (M_{S_2} - M_{S_1})^2\right) \left(\hat s - (M_{S_2} + M_{S_1})^2\right) \right]^{3/2}\,, 
\eeqa
with $Q_u = + 2/3$ as the up-quark electric charge. Folding in the parton distribution function and including all parton contributions, we show the production cross section at the 14 TeV LHC in the left panel of Fig.~\ref{fig:doublelectron}. In this plot, we choose a cut-off $\Lambda_2 = 1$~TeV  and fix the mass difference $M_{S_2} - M_{S_1}=100$~GeV. The differential distribution of the $S_2$ width has the following formula
\beqa
\frac{d\Gamma}{dm_{e^+e^-}^2}
=\frac{e^4}{768\,\pi^3\,\Lambda_2^4\, M_{S_2}^3}
\left\{ [M_{S_2}^2-(M_{S_1}-m_{e^+e^-})^2]
        [M_{S_2}^2-(M_{S_1}+m_{e^+e^-})^2] \right\}^{3\over2}   \,.
\eeqa
In the right panel of Fig.~\ref{fig:doublelectron}, we show the invariant mass distribution of $e^-$ and $e^+$ from the decay of $S_2 \rightarrow S_1\, \gamma^* \rightarrow S_1\,e^- \,e^+$. For this case, the trivial ``dip" appears at $m_{e^- e^+} = 0$ or when the photon invariant mass is zero. 
%

\subsubsection{Case III: one massive gauge boson with two identical massless gauge bosons}
\label{sec:case-III}
For this case, we consider the massive gauge boson to be a SM gauge singlet, $Z^\prime$. There are only two independent operators for the $Z^\prime \to gg$ (or $\gamma\gamma$) vertex:
\beqa
{\cal O}_{Z^\prime GG} = \frac{(\partial^\alpha Z^\prime_\alpha) (G_{\mu\nu} G^{\mu\nu})}{4\,\Lambda_3^2} \,, \qquad
{\cal O}_{Z^\prime G\widetilde{G}} =  \frac{(\partial^\alpha Z^\prime_\alpha) (G_{\mu\nu} \widetilde G^{\mu\nu})} { {4\,\Lambda_4^2} }  \,,
\label{eq:operator-ZpGG}
\eeqa
where $G$ represents either gluon or photon. These operators provide vanishing amplitudes when all three gauge bosons are mass on-shell. If any gauge boson is mass off-shell, a nonzero amplitude exists. Other operators can be shown either vanishing or identical to the two operators in Eq.~(\ref{eq:operator-ZpGG}) up to a total derivative. For instance, one can show $Z^\prime_{\mu\nu} G^{\nu \alpha} \widetilde{G}_{\alpha \mu} = 0$, $Z^{\prime \mu}(\partial_\alpha G_{\mu\beta}) G^{\alpha \beta} = - {\cal O}_{Z^\prime GG}$, $Z^{\prime \mu} (\partial_\alpha G_{\mu\beta}) \widetilde{G}^{\alpha \beta} =  {\cal O}_{Z^\prime G\widetilde{G}}$. For two photon operators, our operator analysis serves as another proof of the Landau-Yang's theorem. 

If the only operators for $Z^\prime$ are these two in Eq.~(\ref{eq:operator-ZpGG}), one could search for the fermion-phobic $Z^\prime$ particle in the dijet final state for $G$ representing a gluon. Similar to the 
operator ${\cal O}_{Z^\prime hh}$ in Section~\ref{sec:case-I}, there is no dip structure from a single $Z^\prime$-mediated dijet invariant mass spectrum. This again depends on how to choose the correct $Z^\prime$ propagator. Using ${\cal O}_{Z^\prime GG}$ as an example, one has the $s$-channel production cross section of $gg \rightarrow Z^\prime \rightarrow gg$ as a function of $\hat{s}$ 
\beqa
\hat{\sigma}(\hat{s}) &=& \frac{1}{512\,\pi\,\Lambda_3^8} \, \frac{\hat{s}^5 \,(\hat{s} - M_{Z^\prime}^2 )^2 }{M_{Z^\prime}^4 \,\left[ (\hat{s} - M_{Z^\prime}^2)^2 + \Gamma_{Z^\prime}^2 M_{Z^\prime}^2 \right] }  
\qquad \mbox{(wrong propagator)}
 \,,   \nonumber \\
\hat{\sigma}(\hat{s}) &=& \frac{1}{512\,\pi\,\Lambda_3^8} \, \frac{\hat{s}^5}{M_{Z^\prime}^4 + \Gamma_{Z^\prime}^2 M_{Z^\prime}^2}  
\qquad \mbox{(right propagator)} \,.
\eeqa
So, there is no dip structure if one chooses the right $Z^\prime$ propagator. The cross section increases dramatically as a function of $\hat s$. This is partially due to the higher-dimensional nature of the effective operator and partially due to the contact interaction mediated by the $Z^\prime$ boson.

After adding the $t$- and $u$-channel contributions and neglecting the width terms, we have the signal cross section as
\beqa
\hat{\sigma}(\hat{s}) \,=\, \frac{1371}{573440\,\pi\,\Lambda_3^8} \frac{ \hat{s}^5 } { M_{Z^\prime}^4 } \,. 
\eeqa
For $\Lambda_3 = 10$~TeV and $M_{Z^\prime} = 500$~GeV, we have the signal production cross section to be around 0.2~fb at the 14 TeV LHC.

Noticing that there exist interference terms between the $Z^\prime$ and QCD contributions, we combine all diagrams together and calculate the differential cross section in terms of the kinematic variable, $\hat{t}$. The pure QCD contribution is proportional to $g_s^4$ and has
\beqa
\frac{d\hat{\sigma}^{\rm QCD}}{d\hat{t}}  = g_s^4\,\frac{9\,(\hat{s}^2 + \hat{s}\hat{t} + \hat{t}^2 )^3 } { 32\,\pi\,\hat{s}^4 \,\hat{t}^2\, (\hat{s} + \hat{t})^2 } \,.
\eeqa
The interference term is proportional to $g_s^2/\Lambda_3^4$ and has its formula as
\beqa
\frac{d\hat{\sigma}^{{\rm QCD}+Z^\prime} }{d\hat{t}} &= &-\frac{g_s^2}{\Lambda_3^4}\,\frac{3 ( 2 \hat s^6 + 6 \hat s^5 \hat t + 15 \hat s^4 \hat t^2 + 20 \hat s^3 \hat t^3 + 15 \hat s^2 \hat t^4 + 6 \hat s \hat t^5 + 2 \hat t^6)}{256\pi\,M_{Z^\prime}^2 \hat s^3 \,\hat t\, (\hat s + \hat t)}\,.
\eeqa
For central jets in the final state with $\hat{t} = - \hat s/2$, the above formula becomes
\beqa
\frac{d\hat{\sigma}^{{\rm QCD}+Z^\prime} }{d\hat{t}} &= & \frac{g_s^2}{\Lambda_3^4}\,\frac{99\,\hat s }{2048\pi\,M_{Z^\prime}^2 } \,. 
\eeqa

If there exists an additional operator for $Z^\prime$ coupling to the SM fermions such as
\beqa
{\cal O}_{Z^\prime t \bar{t}} = g_t \,Z^\prime_\mu \bar{t} \gamma^\mu \gamma_5 t \,,
\eeqa
one could also look for the production of $gg \rightarrow Z^{\prime *} \rightarrow t \bar{t}$ for searching for this fermion-phobic $Z^\prime$. The parton-level production cross section just from the $Z^\prime$ contribution is
\beqa
\hat \sigma(\hat s) &=& \frac{N_c\,g_t^2}{128\,\pi\,\Lambda_3^4}\, \frac{m_t^2}{M_{Z^\prime}^4}
\frac{\hat s^2\,(\hat{s} - M_{Z^\prime}^2)^2 }{(\hat{s} - M_{Z^\prime}^2)^2 + \Gamma_{Z^\prime}^2 M_{Z^\prime}^2 } \sqrt{1 - \frac{4 \,m_t^2 }{\hat s} }   \qquad  \mbox{(wrong propagator)}  \,, \\
\hat \sigma(\hat s) &=& \frac{N_c\,g_t^2}{128\,\pi\,\Lambda_3^4}\, \frac{m_t^2\,\hat s^2}{M_{Z^\prime}^4 + \Gamma_{Z^\prime}^2 M_{Z^\prime}^2} \sqrt{1 - \frac{4 \,m_t^2 }{\hat s} }  \qquad  \mbox{(right propagator)} \,.
\, 
\eeqa
with $N_c=3$ as the color factor. Since there is no interference term from $Z^\prime$ and QCD, one can directly compare the signal and the QCD productions. Assuming that the main decay width of $Z^\prime$ comes from $Z^\prime \rightarrow t \bar t$, we have the width of $Z^\prime$ to be
\beqa
\Gamma_{Z^\prime} = \frac{g_t^2}{4\pi}\,M_{Z^\prime} \left( 1 - \frac{4\,m_t^2}{M_{Z^\prime}^2 } \right)^{3/2} \,, 
\eeqa
which will be used in the later spectrum distribution calculation.

\begin{figure}[th!]
\begin{center}
\includegraphics[width=0.46\textwidth]{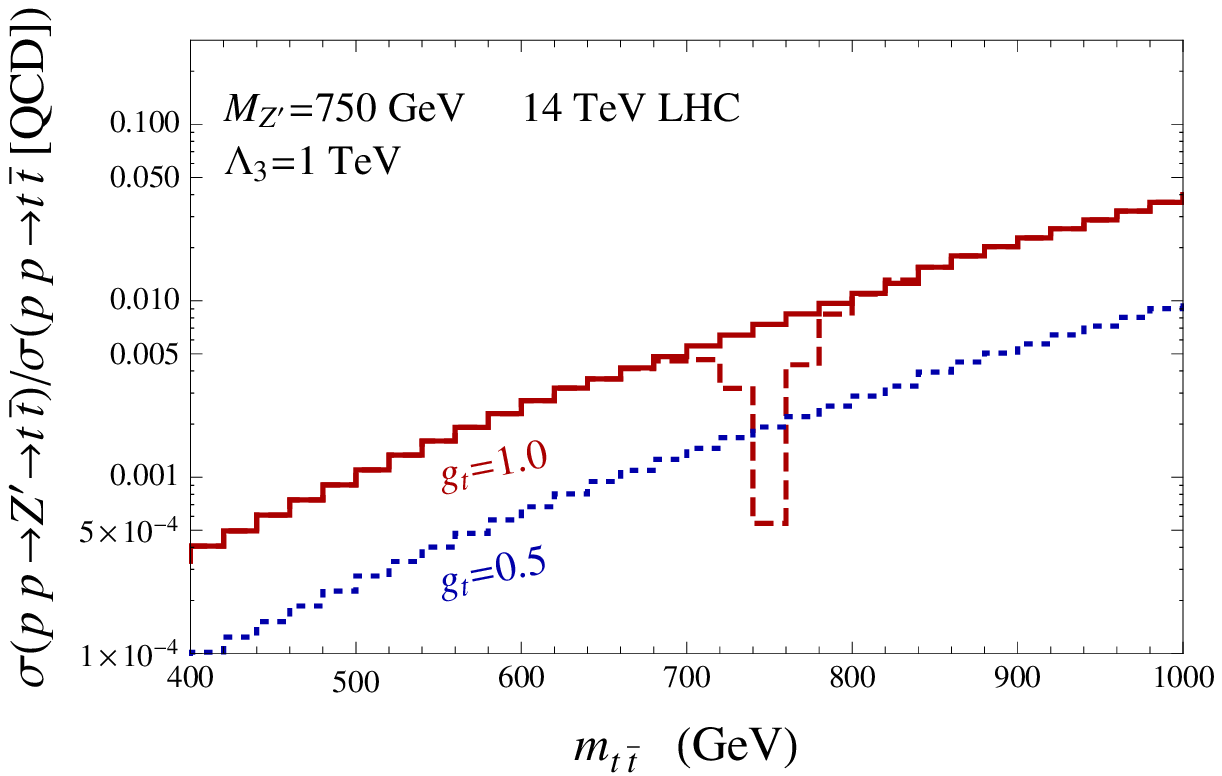} \hspace{3mm}
\includegraphics[width=0.46\textwidth]{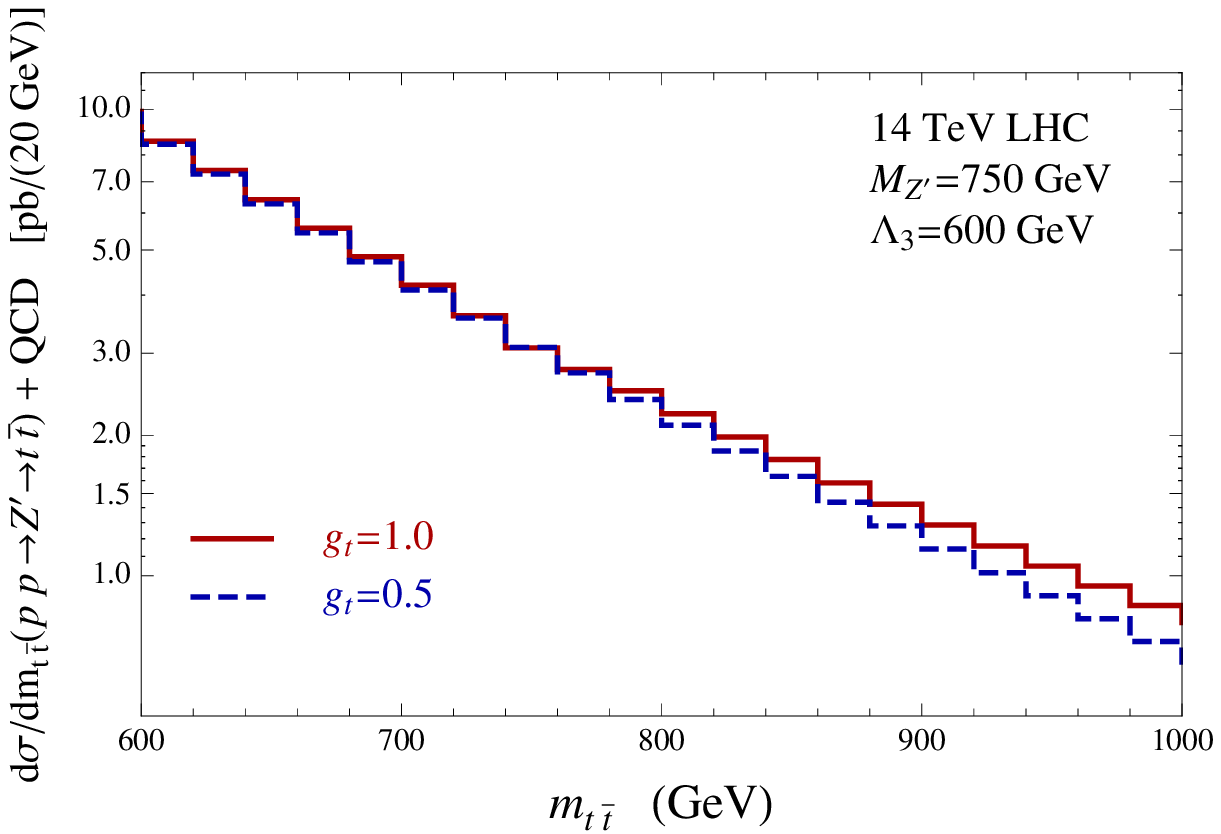} \hspace{3mm}
\caption{Left panel: the ratio of $pp \rightarrow Z^\prime \rightarrow t \bar t$ production cross section over the QCD production as a function of $t\bar t$ invariant mass. The bin size is chosen to be 20 GeV. The red solid and the blue dotted lines have used the correct $Z^\prime$ propagator. The red dashed line has used the wrong $Z^\prime$ propagator and has a dip structure. Right panel: the total production of the signal plus the QCD background as a function of $t\bar t$ invariant masses.}
\label{fig:mttratio}
\end{center}
\end{figure}

In the left panel of Fig.~\ref{fig:mttratio}, we show the distribution of the ratio of the signal production from an off-shell $Z^\prime$ over the QCD tree-level $t \bar t$ production. One can see that the ratio of signal over background is below around 0.1 even for a large coupling $g_t=1.0$ with a cutoff $\Lambda_3 = 1$~TeV. As a comparison to the continuous distributions in the red solid and blue dotted lines, we also show the dip structure in the red dashed line if one uses the wrong $Z^\prime$ propagator. Since the signal production is suppressed by $1/\Lambda_3^4$ from the higher-dimensional nature of this $Z^\prime$ coupling to two gluons, discovering a $Z^\prime$ in this $t\bar t$ channel requires a high luminosity LHC. As an illustration about the total signal plus background spectrum, we choose $\Lambda_3=600$~GeV to enhance the signal production cross section and show the spectrum distribution in the right panel of Fig.~\ref{fig:mttratio}. After added together with the QCD background, the spectrum shows a more dramatical deviation from the SM prediction for a large value of $m_{t \bar t}$. This is similar to the case of a signal production from a higher-dimensional operator including two top quarks.

One can easily extend the simplest $Z^\prime$ case to ``coloron" or ``axi-gluon" models~\cite{Chivukula:1996yr,Simmons:1996fz,Dobrescu:2007yp,Bai:2010dj,Bai:2011mr,Chivukula:2013xka,Chivukula:2014npa}. For three-particle interactions and in contrary to the color-singlet $Z^\prime$ case, there are only two classes of operators using different QCD group structure constants
\beqa
&& {\cal O}^d_{G^\prime G G} = ( D^\mu G^{\prime\, a}_\mu)\, G^b_{\alpha\beta}\, G^{c\,\alpha\beta} \,d_{abc} \,, \qquad
{\cal O}^d_{G^\prime G \tilde{G}}  = ( D^\mu G^{\prime\, a}_\mu) \, G^b_{\alpha\beta}\, \tilde{G}^{c\,\alpha\beta} \,d_{abc} \,,
\nonumber \\
&& {\cal O}^f_{G^\prime G G}  = G_\mu^{\prime\, a} \,( D^\mu G^b_{\alpha\beta})\, G^{c\,\alpha\beta}\, f_{abc} \,, \qquad
{\cal O}^f_{G^\prime G \tilde{G}}   = G_\mu^{\prime\, a} \,( D^\mu G^b_{\alpha\beta})\, \tilde{G}^{c\,\alpha\beta}\, f_{abc}  \,.
\eeqa
The first two charge-conjugation violating operators, ${\cal O}^d_{G^\prime G G}$ and ${\cal O}^d_{G^\prime G \tilde{G}}$, are similar to the $Z^\prime$ case. They provide vanishing on-shell interactions but non-zero off-shell interaction. The last two charge-conjugation conserving operators, ${\cal O}^f_{G^\prime G G}$ and ${\cal O}^f_{G^\prime G \tilde{G}}$, can have gluons with different color indexes and do not satisfy the condition of two identical massless gauge bosons in the Landau-Yang's theorem. The interaction vertex is proportional to $ \epsilon_3 \cdot (p_1 - p_2) \, \epsilon_1 \cdot \epsilon_2$ and $\varepsilon_{\alpha \beta \gamma \sigma} \epsilon^\alpha_3 \epsilon^\beta_1 \epsilon^\gamma_2 ( p_1^\sigma + p_2^\sigma )$, respectively. So, one can have on-shell decay of $G^\prime$ to two massless gluons for the last two operators. 

If only the ``$d_{abc}$" operators exist, the $G^\prime$ can also be produced in pairs from the QCD interaction. It has the on-shell three-body decay $G^\prime \rightarrow ggg, \bar{q}q g, \bar{t} t g$, so one could also look for this ``fermion-phobic coloron" in $6j$, $2t+2\bar{t}+2j$ and $t\bar{t}+4j$ final state if both $G^\prime$'s decay to three objects. Another interesting signature is to have one on-shell $G^\prime$ and the other off-shell $G^\prime$ into two jets. So, other than the three-jet bump from the on-shell $G^\prime$, there is another dip in the remaining two jets. The third interesting signature is to consider the single on-shell $G^\prime$ production in association with one gluon, $gg \rightarrow G^\prime g$. The $G^\prime$ may be discovered as a three-jet bump in the four-jet final state or $t\bar t j$ resonance in the $t\bar t jj$ final state (see the similar situation for the SM $Z$ boson in Ref.~\cite{vanderBij:1988ac}).

\section{Dips from Form Factors}
\label{sec:form-factor}
As we have learned from the no dip results in the vanishing mass-on-shell interaction examples, the vanishing of the vertex happens at the same kinematics location as the complex pole of the massive particle propagator. One way to generate a dip structure is to decouple this correlation and to have the vertex vanish at a different location compared to the massive particle mass. One can realize such a situation by introducing a form factor for the interactions. 

It is well known that the nuclear space-like form factors have dip structures~\cite{Hofstadter:1956qs}. Since we are interested in invariant mass distributions of final state particles at colliders, we need to have dips in the time-like form factors. The simplest example requires some interference effects to have a non-trivial time-like form factor. In this section, we provide one example to have an isolated dip structure. 

For a new SM singlet scalar particle $\phi$, we can have a nontrivial form factor for it to couple to two gluons: $F(q^2) \phi\, G^2_{\mu\nu}$ with $q$ as the momentum of the $\phi$ particle. If the form factor, $F(q^2)$, has a root for some values of time-like $q^2$, we may have a dip in terms of the final particles that $\phi$ couples. To be more concrete, we introduce two QCD color-octet real scalars $S_1 = S_1^a\,T^a$ and $S_2 = S_2^a\,T^a$ to generate the interaction of $\phi$ with two gluons. The Lagrangian is
\beqa
{\cal L} \supset  - \frac{1}{2} M_\phi^2 \phi^2 - M_{S_1}^2 \mbox{Tr} \left[ S_1 S_1\right]- M_{S_2}^2 \mbox{Tr} \left[ S_2 S_2\right] - \mu_1\, \phi\,\mbox{Tr} \left[ S_1 S_1\right] - \mu_2\, \phi\,\mbox{Tr} \left[ S_2 S_2\right]   \,-\lambda_\tau \, \phi \,\overline{\tau} \tau   \,.
\eeqa
Here, we also introduce a coupling of $\phi$ to two $\tau$'s to have a di-tau signature at colliders. From the cubic interaction of $\phi S_i S_i$ and the QCD interactions of $S_i$, we have the following interaction between $\phi$ and two gluons:
\beqa
-F(q^2) \, \frac{\alpha_s}{4\pi}\left( \frac{\mu_1}{M_{S_1}^2} +  \frac{\mu_2}{M_{S_2}^2} \right) \,\frac{ \phi}{4} G^a_{\mu\nu} G^{a\,\mu\nu} \,\equiv\, - \frac{3 \,\alpha_s}{4\pi} \left[ \frac{\mu_1}{M_{S_1}^2} f(\tau_1) + \frac{\mu_2}{M_{S_2}^2} f(\tau_2) \right]\,\frac{ \phi}{4} G^a_{\mu\nu} G^{a\,\mu\nu}   \,,
\label{eq:Leff-phi-GG}
\eeqa
where $\tau_i \equiv q^2/(4 M_{S_i}^2)$ and $F(0)=1$ by our normalization choice. The function $f(\tau)\equiv \tau^{-1} - \tau^{-2}\, h(\tau)$ with 
\beqa
h(\tau) = \left\{ 
\begin{matrix}
\renewcommand{\arraystretch}{1.5}
 \arcsin^2{\sqrt{\tau}} &\quad \tau \le 1  \,, \\
 -\frac{1}{4} \left[ \ln \left( \frac{1 + \sqrt{1 - \tau^{-1}} }{ 1 - \sqrt{1 - \tau^{-1}}  } \right) - i\pi \right]^2  &\quad \tau > 1 \,.  \\
\end{matrix}
\right.   \label{eq:h-function}
\eeqa
For a small value of $\tau$, one has $f(\tau) = 1/3 + 8\,\tau/45$. To have a time-like zero for the form factor, one can have $M_{S_1} < M_{S_2}$, $\mu_1 > 0$, $\mu_2 <0$ and $(M_{S_2}/M_{S_1})^2 < |\mu_2|/\mu_1 < (M_{S_2}/M_{S_1})^4$. In this limit of $q^2 \ll M^2_{S_1} < M^2_{S_2}$,  we have a zero of the form factor at
\beqa
q^2_{\rm zero} = \frac{15 \, M_{S_2}^2}{2} \, \frac{|\mu_2|/\mu_1 - M_{S_2}^2/M_{S_1}^2 }{  M_{S_2}^4/M_{S_1}^4 - |\mu_2|/\mu_1    } \,. 
\eeqa

For a very heavy $\phi$ field beyond the collider parton center-of-mass energy, we can integrate out the $\phi$ field and have the following effective interaction 
\beqa
F(\hat s) \, \frac{\alpha_s}{4\pi \,\Lambda^3}\,\frac{1}{4}\,G^a_{\mu\nu} G^{a\,\mu\nu} \,\overline{\tau} \tau \equiv F(\hat s) \, \frac{\alpha_s}{4\pi}\left( \frac{\mu_1}{M_{S_1}^2} +  \frac{\mu_2}{M_{S_2}^2} \right) \,\frac{\lambda_\tau}{4\,M_\phi^2} G^a_{\mu\nu} G^{a\,\mu\nu} \,\overline{\tau} \tau \,, 
\eeqa
where $\hat s$ is the invariant mass of the two $\tau$'s in the final state. The production cross section in the center-of-mass frame is 
\beqa
\hat\sigma(\hat s) = \frac{\alpha_s^2 \, {\hat s}^2}{8192 \pi^3 \Lambda^6} \, |F(\hat s)|^2 \,.
\eeqa
\begin{figure}[th!]
\begin{center}
\includegraphics[width=0.5\textwidth]{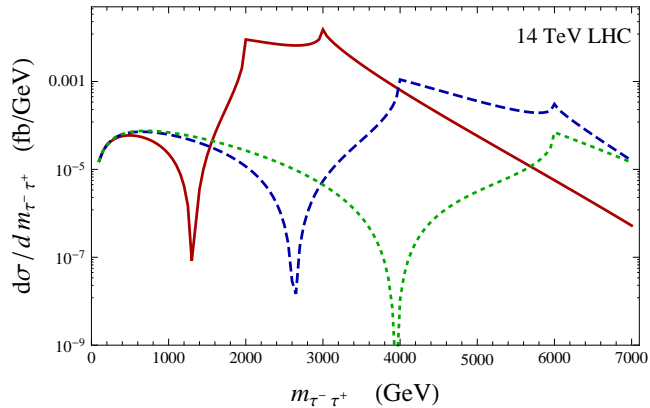}
\caption{The signal differential cross section as a function of the $\tau^- \tau^+$ invariant masses. The cutoffs for the three curves are fixed to be $\Lambda =1$~TeV. From left to right, the three dips correspond to $M_{S_1} = 1, 2, 3$~TeV with $M_{S_2}=1.5 M_{S_1}$ and $\mu_2/\mu_1=- 8/3$.}
\label{fig:form-factor-scalar}
\end{center}
\end{figure}

After folding in the parton distribution function (PDF), we show the differential production cross sections in Fig.~\ref{fig:form-factor-scalar} at the 14 TeV LHC. As we increase the particle masses responsible for the form factor, the dip locations move to a higher value. Because the loop-generated function in Eq.~(\ref{eq:h-function}) has a bump feature at $2M_{S_1}$ and $2M_{S_2}$, the cross sections also show two bumps. The relative heights of the bumps become lower and lower as we increase $M_{S_1}$ and $M_{S_2}$. This is simply due to the reduction of PDF's at a large value of center-of-mass energy. Because the effective operator is loop-generated, it generically predicts a very small rate at the 14 TeV LHC.

\section{Dip Plus Bump Structure}
\label{sec:dip-bump}
In this section, we also discuss the ``standard way" to obtain a dip structure that coexists with a bump. A new particle can interfere with the SM background matrix elements and generate a dip plus bump kinematic distribution. Using the production of $u\bar u \rightarrow \mu^+\mu^-$ from $\gamma^*$, $Z$ and $Z^\prime$ as an example, we have the production cross section as~\cite{Barger:1980ix} 
\beqa
\hat{\sigma}(u\bar{u} \rightarrow \mu\bar{\mu}) = \frac{\hat{s}}{72\,\pi} \left(|H_-|^2 + |H_+|^2 + |H_-^\prime|^2 + |H_+^\prime|^2 \right)  \,,
\eeqa
with the four different helicity contributions as
\beqa
H_{\pm} = -\frac{Q_u \,e^2}{\hat s} + \sum_{i = Z, Z^\prime} \frac{g^i_V(\mu) g^i_V(u) \pm g^i_A(\mu)g^i_A(u) }{\hat s - M_i^2 + i\,M_i \Gamma_i}  \,,\quad
H_\pm^\prime = \sum_{i = Z, Z^\prime} \frac{g^i_V(\mu) g^i_A(u) \pm g^i_V(u)g^i_A(\mu) }{\hat s - M_i^2 + i\,M_i \Gamma_i} \,.
\eeqa
Here, the couplings are defined as ${\cal L} \supset -[g^i_V(\psi) \bar{\psi} \gamma_\mu \psi + g^i_A(\psi) \bar{\psi} \gamma_\mu \gamma_5 \psi]Z^\mu_i$. We have also neglected the mixing between $Z$ and $Z^\prime$. For a sequential $Z^\prime$ with couplings proportional to the $Z$ boson couplings, $g^{Z^\prime}_{V, A}(\mu) = \sqrt{\xi}\, g^Z_{V, A}(\mu)$,
we have a dip in the production cross section at $\sqrt{\hat s} = M_{Z^\prime}/\sqrt{1+\xi}$ for $M_Z \ll M_{Z^\prime}$. 

\begin{figure}[th!]
\begin{center}
\includegraphics[width=0.46\textwidth]{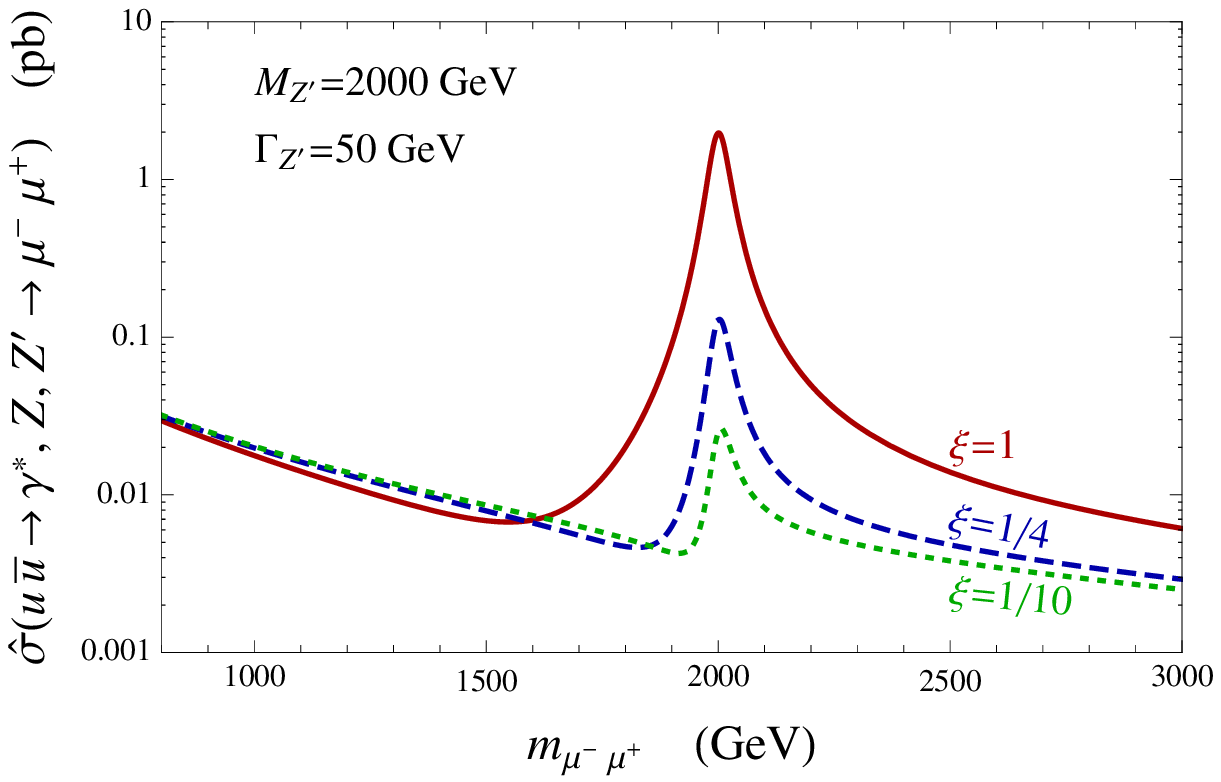} \hspace{3mm}
\includegraphics[width=0.46\textwidth]{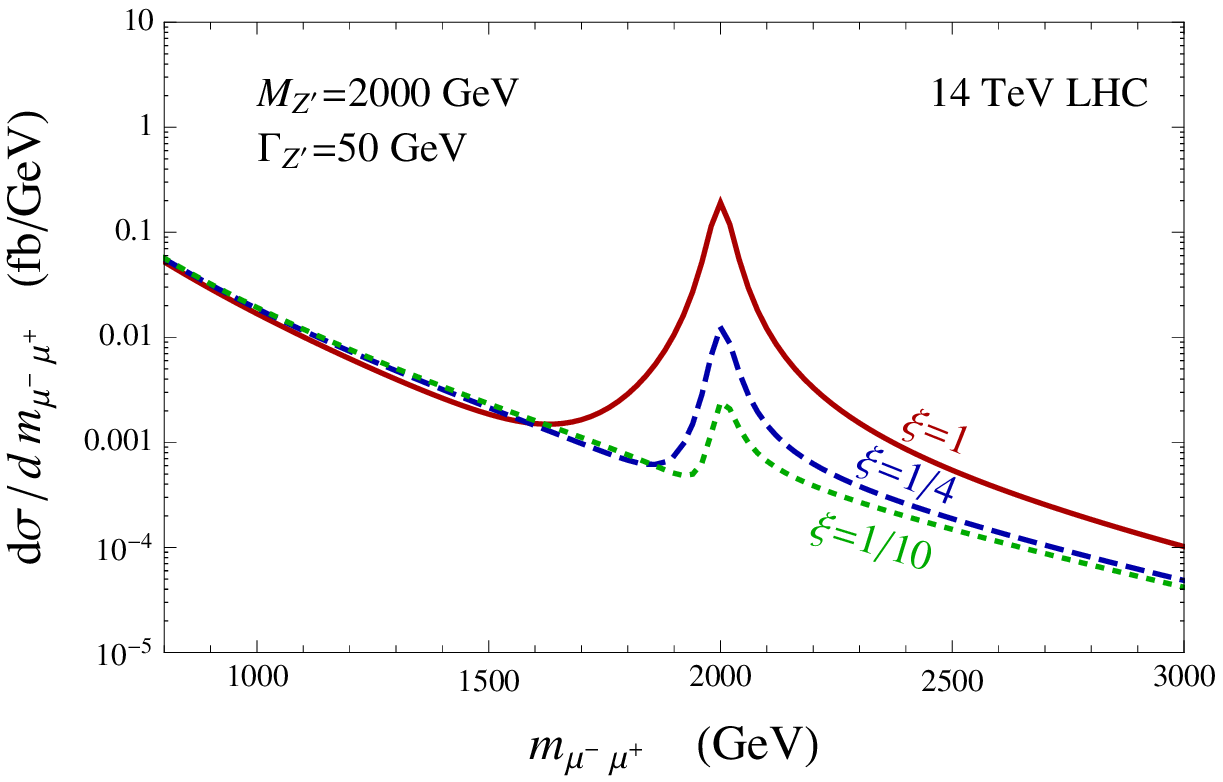} \hspace{3mm}
\caption{Left panel: the parton-level production cross section of two muons from Drell-Yan processes including $Z^\prime$. The parameter $\xi$ measures the couplings of $Z^\prime$ to fermions with respect to the $Z$ boson couplings. Right panel: the production cross section as a function of $m_{\mu^- \mu^+}$ at the 14 TeV LHC.}
\label{fig:sigmamumu}
\end{center}
\end{figure}

In Fig.~\ref{fig:sigmamumu}, we show the differential production cross section of two muons from the Drell-Yan processes at the 14 TeV LHC. As one can clearly see, a dip structure before a bump can exist with the suitable relations between the $Z^\prime$ and $Z$ couplings (see Ref.~\cite{Accomando:2013sfa} for a recent study). The standard ``bump search" should be modified to find the ``dip+bump" signal structure. From the left panel of Fig.~\ref{fig:sigmamumu}, one see that for a smaller value of $\xi$ the locations of dip and bump become closer and both heights become smaller. A similar dip plus bump structure appears in Ref.\cite{Ralston:2012ye}, where  a concern of 
the experimental Higgs analysis at 125 GeV was raised.

\section{Discussion and Conclusions}
\label{sec:conclusion}
We have investigated the signature of a dip in the invariant mass distribution of the final products of the high energy processes. 

First, we systematically look at the vanishing decay amplitudes, which can play an important role in exploring the
fundamental selection rule of the new particle. For example, it may give us useful information of the spin and the QCD color of the new object. We have analyzed various cases that the decay amplitudes vanish when all particles in the relevant vertex are mass on-shell. The overall amplitude also involves the propagator of the exchanged unstable particle with a complex mass pole. Unfortunately, a consistent use of the complex mass in the mass parameter results in the ``no-dip" overall amplitude even though the decay amplitude is zero. Nevertheless, we have identified interesting collider signatures for the fermion-phobic $Z^\prime$ or $G^\prime$, which naturally behaves a three-body resonance.

We then explore another way to generate a dip by constructing a zero in the  vertex form factor of the time-like momentum transfer. We illustrate this scenario by a new scalar that couples to two gluons with subtle physics at a higher energy scale. Finally, we also point out the interesting ``dip+bump" signature from the ``standard way'' of destructive interference among various resonances. If the bump can not be identified by itself due to a weak signal strength, one could look for a nearby deficit of events to improve the reach for a ``dip+bump" signature. A new customized dip+bump function together with a continuous background function should be developed to fit the data in a similar manner as in Ref.~\cite{CMS:kxa,Aad:2014aqa}. 

Usually, the dip from new physics is hidden in a large background from the SM processes. Identifying a dip structure may require
high luminosity colliders, as well as new algorithms in searching for
the depletion of events around the new particle mass. It is our
wish that this article opens our vision of how new physics can appear
very different from the conventional search of bumps in the mass
distributions.

\subsection*{Acknowledgements}
We would like to thank Nima Arkani-Hamed, Vernon Barger, Bogdan Dobrescu, David Kosower and Ian Low for useful discussion. YB is supported by the U. S. Department of Energy under the contract DE-FG-02-95ER40896. WYK is supported by the U. S. Department of Energy under the contract DE-FG-02-12ER41811. YB thanks the Center for Future High Energy Physics and the Aspen Center for Physics, under NSF Grant No. PHY-1066293, where parts of this work are finished.

\providecommand{\href}[2]{#2}\begingroup\raggedright\endgroup

 \end{document}